\documentstyle[11pt,newpasp,twoside]{article}
\def\edcomment#1{\iffalse\marginpar{\raggedright\sl#1\/}\else\relax\fi}
\reversemarginpar

\begin{document}
\title{VLBA Observations and Effelsberg Monitoring of a Flaring Water Maser in Mrk~348}
 \author{A.~B.~Peck, H.~Falcke, C.~Henkel, K.~M.~Menten}
\affil{Max-Planck-Institut f\"ur Radioastronomie, Auf dem H\"ugel 69, D-53121 Bonn, Germany}

\begin{abstract}
We report on single-dish monitoring as well as extremely high angular
resolution observations of the flaring H$_2$O maser in the Seyfert 2
galaxy Mrk 348.  The H$_2$O line is redshifted by $\sim$130 km
s$^{-1}$ with respect to the systemic velocity, is very broad, with a
FWHM of 130 km s$^{-1}$, and has no detectable high velocity
components within 1500 km s$^{-1}$ on either side of the strong line.
Monitoring observations made with the Effelsberg 100m telescope show
that the maser varies significantly on timescales as short as one day.
VLBA observations indicate that the maser emission arises entirely
from a region less than 0.25 pc in extent, located toward a continuum
component thought to be associated with the receding jet.

\end{abstract}

\section{Introduction}

Mrk~348 (NGC 262) is a Seyfert 2 galaxy at a redshift of 0.01503
(Huchra \mbox{\it et al.} 1999).  The galaxy is classified as an S0
with a low inclination, and exhibits a large H\kern0.1em{\sc i} halo
which may have been produced by an interaction with the companion
galaxy NGC 266 (e.g. Simkin \mbox{\it et al.} 1987).  VLBA images
(Ulvestad \mbox{\it et al.} 1999) reveal a small-scale double
continuum source, the axis of which is aligned with the optical
([OIII], Capetti \mbox{\it et al.} 1996) emission.  Astrometry
indicates that the optical and VLBI cores are nearly coincident.
Ground-based observations (Simpson \mbox{\it et al.} 1996) show
evidence of a dust lane crossing the nucleus and an ionization cone.
Attempts to detect the expected obscuring torus at radio wavelengths
(e.g. H\kern0.1em{\sc i}, Gallimore \mbox{\it et al} 1999; CO,
Taniguchi \mbox{\it et al.} 1990; free-free absorption, Barvainis \&
Lonsdale 1998) have not been successful.

The compact radio source in Mrk~348 is unique among Seyferts in that
it is very bright and extremely variable.  The observations presented
here were made during a local minimum, when the total continuum flux
density at 22 GHz was $\sim$0.6 Jy.

\begin{figure}
\plotfiddle{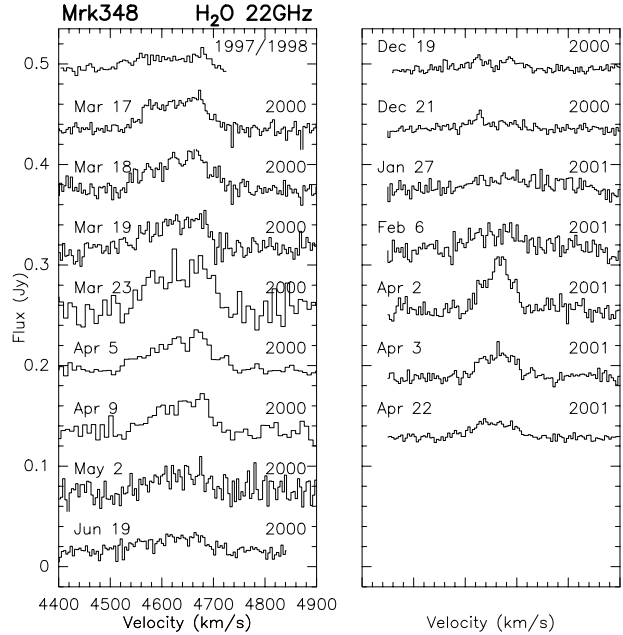}{7.5cm}{-90}{43}{43}{-120}{240}
\caption{Single dish profiles from Effelsberg 100m telescope.  The
peak flux in the line was $\sim$40 mJy on April 9, but decreased to 18
mJy by June 19.  In April 2001, the line peak again increased to 50 mJy, but the FWHM decreased to $\sim$65 km s$^{-1}$.  This latest flare lasted only a few days.}
\end{figure}

\section{Observations}

The initial detection of the flaring maser in Mrk 348 using the
Effelsberg 100m telescope took place in 2000 March (see Fig.~1, 2nd
profile).  Re-analysis of previous data on this source (shown in the
top profile, Fig.~1) indicates that the maser was also present but
only marginally detectable in late 1997.  The H$_2$O maser line in Mrk
348 is extremely broad, with a FWHM of $\sim$130 km s$^{-1}$, though
in many of the monitoring epochs the emission appears to consist of 2
lines which can be fit by Gaussian functions with FWHM of $\sim$60 km
s$^{-1}$ and $\sim$100 km s$^{-1}$ respectively, separated by $\sim$70
km s$^{-1}$.  There are no detectable high velocity components within
1500 km s$^{-1}$ on either side of the strong emission line.

Monitoring through June 2000 showed that the maser again decreased to
its original level within 2 months.  The June profile (bottom profile,
Fig.~1) indicates a peak flux of $\sim$18 mJy.  This is consistent
with the VLBA results shown in Fig.~2, taken on June 10, 2000.  The
monitoring observations indicated that the line was too broad
(FWZP$>$250 km s$^{-1}$) to fit in a single 16 MHz VLBA IF, so 2 IFs
of 16 MHz each were used, overlapped by 5 MHz.  Following calibration,
the overlapping channels were removed and the 2 IFs were added
together to yield a single cube of 174 channels covering 23 MHz.  Line
profiles of the resulting cube are shown in Fig.~2, superimposed on a
continuum map made from 20 line-free channels. The maser emission is
clearly seen to lie along the line of sight to the jet, rather than
the core which is thought to lie nearer the bright component of the
continuum source.  No maser emission is seen toward any other region
of the radio source, only toward the jet, and here the maser emission
is unresolved at our angular resolution of 0.42$\times$0.76 mas.  This
corresponds to a linear size of less than 0.25 pc (assuming H$_0$=75
km s$^{-1}$Mpc$^{-1}$).  The Gaussian fit to the line shown in the
first profile has an amplitude of 16$\pm$2 mJy and an integrated flux
of 2.4$\pm$0.3 Jy/beam/km s$^{-1}$, indicating that all of the flux
measured in the Effelsberg 19 June observation has been recovered.
The FWHM is 142$\pm$9 km s$^{-1}$ centered on V$_{\rm LSR}$=4640$\pm$2
km s$^{-1}$, consistent with the single-dish measurements and
redshifted by 131 km s$^{-1}$ with respect to the systemic velocity.
This redshift might indicate that it is the receding jet that impacts
the molecular cloud, and that the masing gas is being entrained.  A
tentative 2 component fit to the H$_2$O profile yields a narrower line
at 4683 km s$^{-1}$ with FWHM $\sim$60 km s$^{-1}$ and amplitude
$\sim$8 mJy, and a broader line at 4620 km s$^{-1}$ with FWHM
$\sim$100 km s$^{-1}$ and amplitude $\sim$12 mJy, again consistent
with our single-dish measurements.

\begin{figure}
\plotfiddle{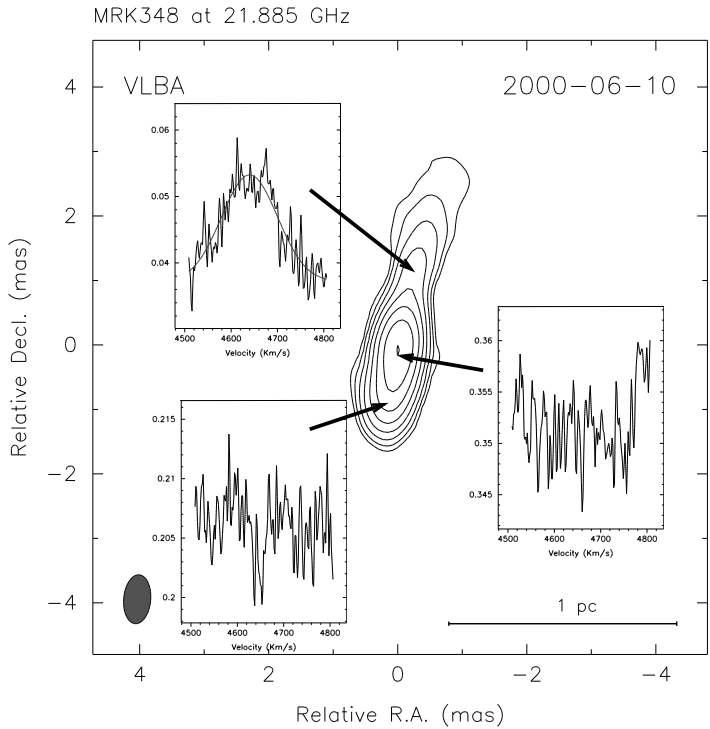}{9.5cm}{0}{50}{50}{-170}{-65}
\caption{Line profiles from VLBI data.  The continuum map is naturally
weighted.  The lowest contour is 3 mJy and the peak is 352 mJy.  The
RMS noise in the continuum map is $<$1 mJy/beam, and in the line
profiles is $\sim$4 mJy/beam/channel.}
\end{figure}

Resumption of the monitoring program at Effelsberg in December 2001
showed little change in the line flux, then in February another flare
appears to have begun.  The observations made in early April 2001 (see
Fig.~1) show that the FWHM in the line decreased to 65 km s$^{-1}$,
and the centroid increased from 4642 km s$^{-1}$ to $\sim$4665 km
s$^{-1}$.  Thus it seems likely that only one of the two line
components was flaring.  Following this peak on 2 April, the line
again began to decrease in flux.  We are continuing our single-dish
monitoring program at the Effelsberg telescope, and will attempt to
observe with the VLBA during the next maser flare.

\section{Conclusions}

During early 2000, the H$_2$O emission toward Mrk~348 showed a
dramatic intensity increase which coincided with a significant
increase in the flux of the nuclear radio continuum source.  The
unusual line profile leads us to suspect that this source, and
possibly NGC~1052, might belong to a class of megamaser galaxies in
which the amplified emission is the result of an interaction between
the radio jet and an encroaching molecular cloud, rather than
occurring in a circumnuclear disk.  Analysis of our recent VLBA
observations indicates that the emission does indeed arise along the
line of sight to the jet in Mrk~348 (see Fig.~2), confirming this
prediction.  The very high linewidth occurring on such small spatial
scales indicates that the H$_2$O emission arises from a shocked region
at the interface between the energetic jet material and the molecular
gas in the cloud where the jet is boring through.  This hypothesis is
supported by the spectral evolution of the continuum source
(Brunthaler, priv. comm.), which showed an inverted radio spectrum
with a peak at 22 GHz, later shifting to lower frequencies. By analogy
to IIIZw2 (Brunthaler \mbox{\it et al.} 2000) this would indicate the
formation and evolution of very compact hotspots propagating through a
dense medium.  In this scenario, the recent high frequency radio
continuum flare and the apparent northward movement of the brightest
continuum component are attributable to the impact between the jet and
the molecular cloud.  This is consistent with recent VLBA continuum
monitoring observations that show that the continuum source does not
appear to be expanding (Ulvestad, {\it priv.comm.}), as one would
expect from a hotspot or working surface.  Rather the jet component
toward which the maser appears might be associated with a stationary
shock.  The very close temporal correlation between the flaring
activity in the maser emission and the continuum flare further suggest
that the masing region and the continuum hotspot are nearly coincident
and may be different manifestations of the same dynamical events.


\begin{references}

{\small
\reference Barvainis, R. \& Lonsdale, C. 1998, \aj, 115, 885
\reference Brunthaler, A., Falcke, H., Bower, G.~C., Aller, M.~F., Aller, H.~D.,Ter\"asranta, H., Lobanov, A.~P., Krichbaum, T.~P. \& Patnaik, A.~R. 2000, \aap, 357, L45
\reference Capetti, A., Axon, D.~J., Macchetto, F., Sparks, W.~B. \& Boksenberg, A. 1996, \apj, 469, 554
\reference Falcke, H. Henkel, C., Peck, A.~B., Hagiwara, Y., Prieto, M.~A. \&
 Gallimore, J.~F. 2000, \aap, 358, L17 
\reference Gallimore, J.~F., Baum, S.~A., O'Dea C.~P., Pedlar, A. \& Brinks, E. 1999, \apj, 524, 684
\reference Huchra, J.~P., Vogeley, M.~S. \& Geller, M.~J. 1999, \apjs, 121, 287
\reference Simkin, S.~M, Su, H.-J., van Gorkom, J. \& Hibbard, J.  1987, Science, 235, 1367
\reference Simpson, C., Mulchaey, J.~S., Wilson, A.~S., Ward, M.~J. \& Alonso-Herrero, A. 1996, \apj, 457, L19
\reference Taniguchi, Y., Kameya, O., Nakai, N. \& Kawara, N. 1990, \apj, 358, 132 
\reference Ulvestad, J.~S., Wrobel, J.~M., Roy, A.~L., Wilson, A.~S., Falcke, H \& Krichbaum, T.~P. 1999, \apj, 517, L81
}

\end{references}
\end{document}